\documentclass{PoS}

\title{The Luminosity function of Narrow-Line Seyfert galaxies based on SDSS DR7 data}

\ShortTitle{The NLSy1 luminosity function}

\author{\speaker{Andrey Ermash}\\
        Astro Space Centre, P.N. Lebedev Physical Institute of Russian Academy of Sciences, Moscow, Russia\\
        E-mail: \email{aermash@gmail.com}}

\abstract{We present measurements of AGN type 1 luminosity function in the forbidden line [OIII]5007\AA~using data from SDSS DR7. 
A special attention is paid to NLSy1. 
A new approach in calculating the luminosity function is present. 
We also account for the large-scale structure variations of the Universe density. 
The results obtained are compared with ones from the literature. 
A prediction of X-ray luminosity function based on our results shows an agreement with observations. 
One of our preliminary conclusions is that NLSy1 seems to occupy a more narrow range in the nuclear luminosity than BLSy1, but the average values are within errors.
}

\FullConference{Nuclei of Seyfert galaxies and QSOs - Central engine \& conditions of star formation - seyfert2012,\\
		November 6-8, 2012\\
		Max-Planck-Insitut f\"ur Radioastronomie (MPIfR), Bonn, Germany}

\begin{document}

\section{Measuring the luminosity function}
At first, we select objects with ${\rm FWHM}(H_\alpha)\geq1200{\rm km s^{-1}}$ (BL~AGN) from SDSS DR7 data.
For our work we use galaxies in the distance interval $d_c=55-738$Mpc.
The lower distance limit is chosen to avoid the influence of the Local Supercluster \cite{tempel11}.
The upper limit ($z=0.18$) is chosen due to impossibility to use our normalization algorithm on larger distances.
The main sample is divided into two subsamples, NLS ($1200{\rm km s^{-1}}\geq{\rm FWHM}(H_\alpha)<2000{\rm km s^{-1}}$, 2082 objects) and BLS (${\rm FWHM}(H_\alpha)\leq2000{\rm km s^{-1}}$, 6938 objects).
 
In order to evaluate AGN luminosity we used forbidden oxygen line [OIII]$\lambda$5007\AA.
It is believed that the contribution from star formation to this line is negligible \cite{haostrauss05}.
Since the emitting gas is located in NLR, the AGN luminosity estimates based on this line do not suffer from the orientation effects.
 
We use a modified $V/V_{max}$ method.
For each luminosity bin we calculate an average probability of detection as a function of comoving distance $\rho(d_c)$.
Knowing this function allows us to estimate $V_{max}$: 
$$\langle V_{max}\rangle=\int_{d_{c,min}}^{d_{c,max}}4\pi r^2\rho(d_c)dr\,,$$
where $d_{c,min}$ and $d_{c,max}$~--- lower and upper distance limits of our sample.
 
When calculating luminosity function we also took into account variations of density of galaxies per unit of volume due to the large-scale structure.
Two facts make it possible: at first, SDSS is complete in a wide range of magnitudes.
Secondly, the luminosity function of inactive galaxies is well known in the Local Universe.
 
Moreover, the normalized amount of AGNs in some volume element is:
$$N_{AGN,norm}=N_{AGN}\frac{\langle\rho_{gal}\rangle}{\rho_{gal}}\,,$$
where $N_{AGN}$ is an observed number of AGN, $\langle\rho_{gal}\rangle$ is an average (calculated from luminosity function) density of galaxies, $\rho_{gal}$~--- the density of inactive galaxies which is actually observed.
It is necessary to consider only galaxies with magnitudes in the completeness interval of the SDSS. 
 
With help of the luminosity function of inactive galaxies, one obtains:
$$\frac{\langle\rho_{gal}\rangle}{\rho_{gal}}=\frac{L_{sch} V}{L_{obs}}=\kappa\,,$$
where $L_{sch}$ is an integral luminosity, calculated using the Schechter function, $L_{obs}$~--- the actually observed total luminosity of galaxies in the considered volume $V$ and luminosity interval.
 
The transition from $L$ to $\rho$ is possible only when the parameters $\alpha$ and $L^*$ of the Schechter function are fixed. 
It is the case for galaxies in the Local Universe . 
 
\begin{figure}
 \includegraphics[angle=0,width=0.42\linewidth,height=0.36\linewidth]{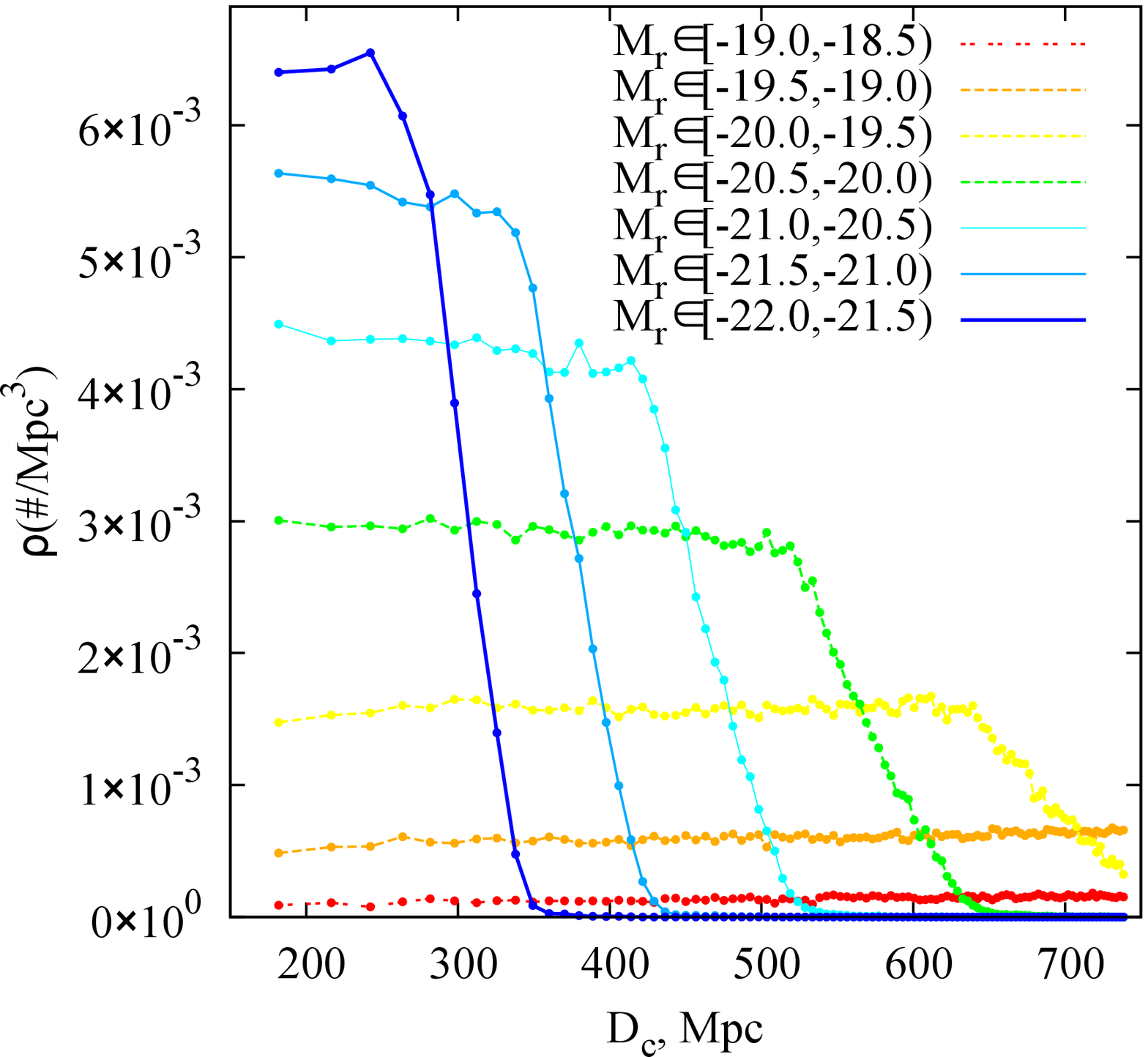}\hfill
 \includegraphics[angle=0,width=0.57\linewidth,height=0.36\linewidth]{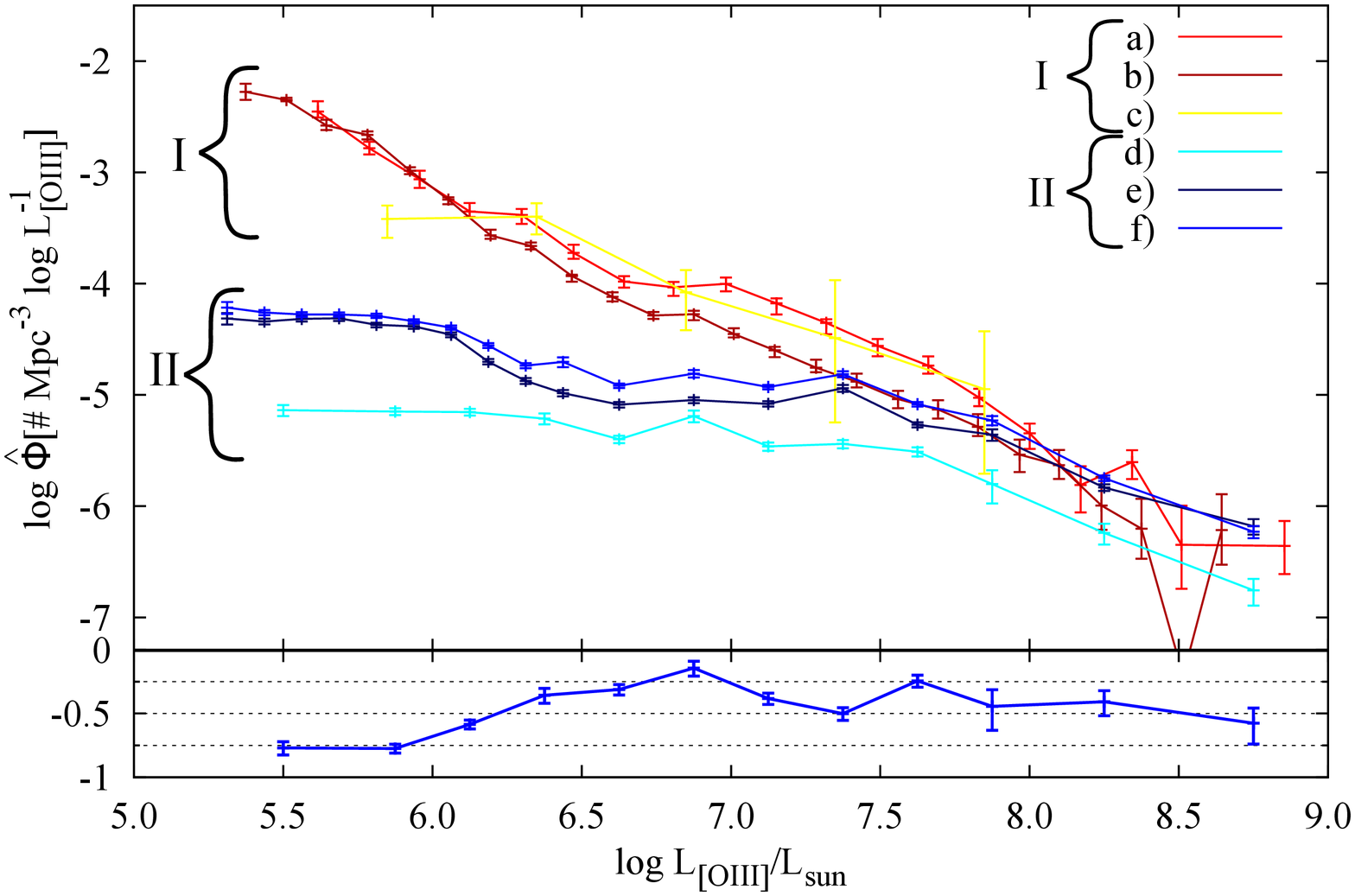}
 \caption{\emph{(LEFT)} The result obtained by applying our normalization to a sample of normal galaxies split into bins by absolute magnitude. \emph{(RIGHT)}
 Luminosity functions calculated in [OIII]5007\AA. 
 a) Sy1 \cite{haostrauss05} 
 b) Sy2 \cite{haostrauss05} 
 c) Type2 AGN \cite{bongiorno10} 
 d,e,f) NLSy1,BLSy1,NLSy1+BLSy1 - this work.
 }
 \label{fig1}
\end{figure}

In fig.~\ref{fig1}\emph{(left)} we show the result obtained by applying this normalization to a sample of normal galaxies split into bins by absolute magnitude.
A plateau followed by a rapid decline when visual magnitude reaches the completeness limit of SDSS survey is clearly seen.
This is exactly the expected behaviour, so our normalization is adequate. 

\begin{figure}
 \includegraphics[angle=0,width=0.49\linewidth,height=0.36\linewidth]{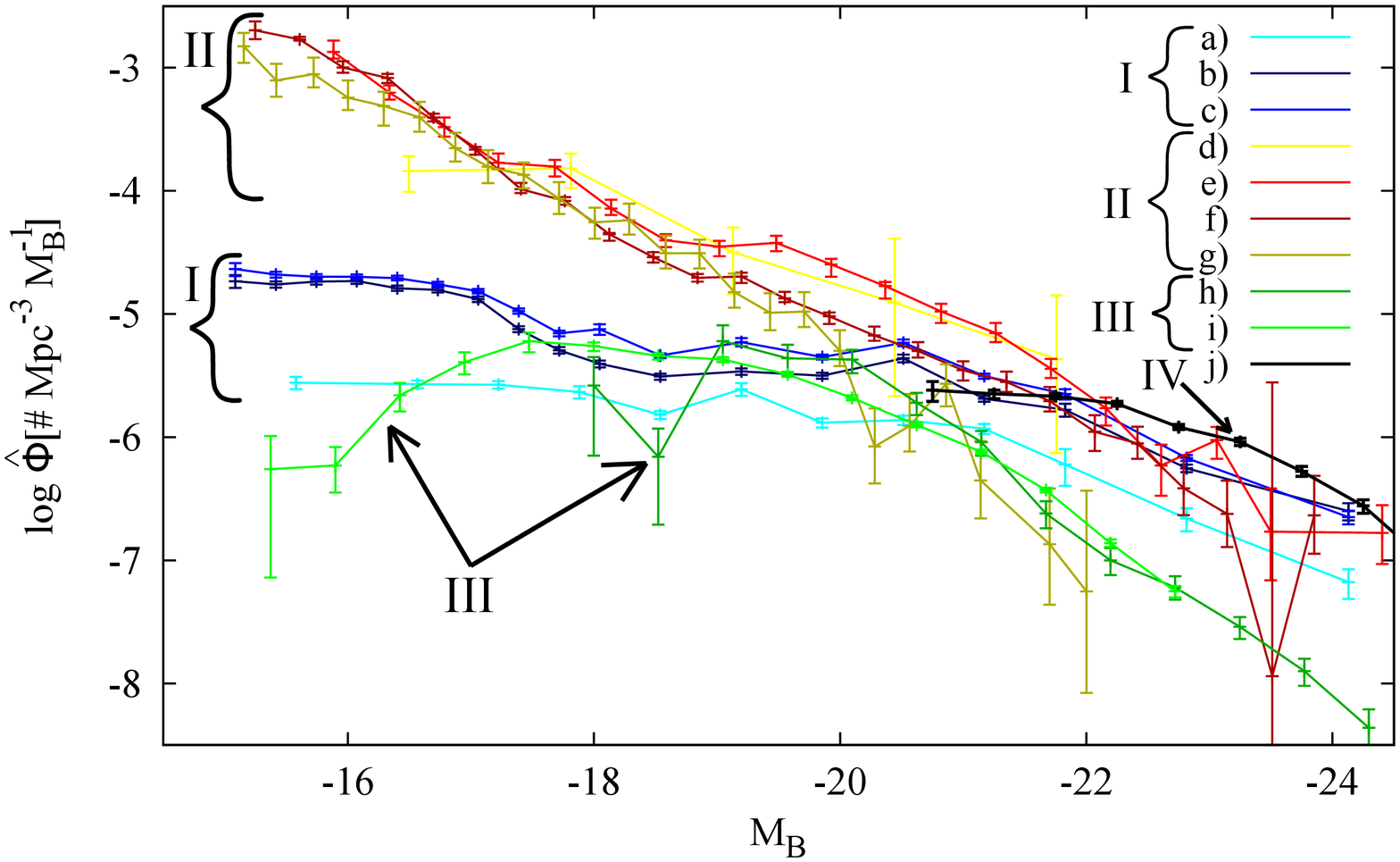}\hfill
 \includegraphics[angle=0,width=0.49\linewidth,height=0.36\linewidth]{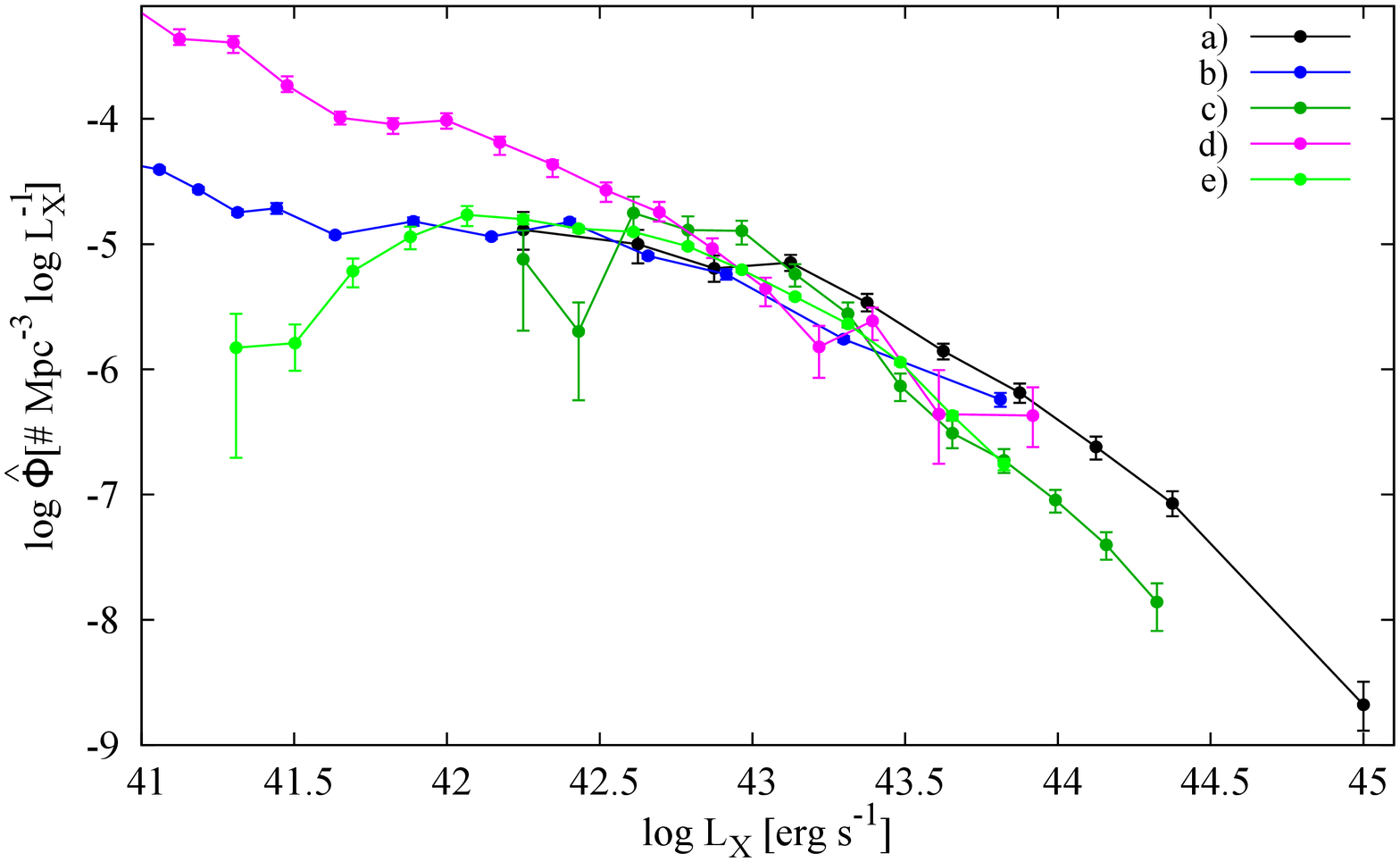}
 \caption
  {\emph{(LEFT)} Luminosity functions in B band obtained by converting from [OIII] or $H_\alpha$ LFs.
  {\bf Group 1.} (a,b,c) NLSy1, BLSy1, NLSy1+BLSy1 converted from [OIII], this work.
  {\bf Group 2.} d) Type2 AGN, [OIII] \cite{bongiorno10} 
                 e) Sy1, [OIII] \cite{haostrauss05}
                 f) Sy2, [OIII] \cite{haostrauss05}
                 g) Sy1+Sy2, $H_\alpha$ \cite{haostrauss05}
  {\bf Group 3.} h) Sy1, $H_\alpha$ \cite{schulze09}
                 i) Sy1, $H_\alpha$ \cite{greeneho09} 
  {\bf Group 4.} g) QSO calculated in the B band \cite{croom04}  
  \emph{(RIGHT)} 
  a) The observed Sy1 X-ray luminosity function (0.5--2kev) \cite{hasinger05}. 
  The predicted Sy1 X-ray luminosity functions, based on 
  b) Our [OIII] LF.
  c) [OIII] LF from \cite{haostrauss05}.
  d) $H_\alpha$ LF from \cite{schulze09}.
  e) $H_\alpha$ LF from \cite{greeneho09}.
  }
 \label{fig2}
\end{figure}
 
\section{The luminosity function}

The obtained luminosity functions of NLS, BLS and Sy1 (NLS and BLS together) are shown in fig.~\ref{fig1}\emph{(right)}.

For a comparison we also plot the luminosity functions from \cite{haostrauss05} and \cite{bongiorno10} calculated in the same spectral line.
Note that in \cite{haostrauss05} LF is obtained for Sy1 and Sy2, while in \cite{bongiorno10} only for Sy2.
 
As can be seen in fig.~\ref{fig1}\emph{(right)}, the results of these two papers are in an agreement with each other. Our luminosity function is significantly lower.
 
In order to compare our results with as many as possible LFs calculated by different authors we convert our LF from [OIII]$\lambda$5007\AA~to the Johnson band B according to \cite{bongiorno10}.
 
The result is shown in fig.~\ref{fig2}\emph{(left)} where the following LFs are also plotted:
\begin{enumerate}
 \item Converted from [OIII] to $M_B$ (\cite{bongiorno10,haostrauss05} and this study)
 \item Converted from $H_\alpha$ \cite{haostrauss05,schulze09,greeneho07}
 \item The quasar luminosity function at redshift $0.4<z<0.88$ measured in B band \cite{croom04}.
\end{enumerate} 
 
As can be seen from these figs LFs obtained by different authors form two distinct groups (groups II and III on fig.~\ref{fig2}\emph{(left)}).
The first and the second ones consist of the results of \cite{haostrauss05,bongiorno10} and \cite{schulze09,greeneho07}, respectfully.
LF of Sy1 obtained by us is between these contradicting results.
It is important that our LF agrees well with the LF of quasars \cite{croom04} on highest luminosities. 
 
It is remarkable that $\phi(NLS)/\phi(BLS)$ is not constant.
Instead it is actually a function of the luminosity (fig.~\ref{fig1}\emph{(right)}., bottom panel).
It has a maximum between $\log{(L_{[OIII]}/L_\odot)}=6.8$ and $7.6$.
At larger luminosities it slightly decreases.
The decrease towards lower luminosities is $\sim0.5dex$ to $\log{(L_{OIII}/L_\odot)}=5.5$.
 
A putative explanation can be proposed if the results of \cite{xukomossa12} are considered.
Authors studied samples of NLS and BLS selected from SDSS DR3.
Ones of the key results of their work are the distributions of $\log{(L/L_{edd})}$ and $\log{(M_{BH}/M_\odot)}$.
The normality of this distributions implies that $\log{(L_{bol})}$ also has a normal distribution with the following parameters:
\begin{itemize}
 \item NLS:$\mu=44.589\pm0.052$, $\sigma=0.404\pm0.043$
 \item BLS:$\mu=44.535\pm0.088$, $\sigma=0.560\pm0.061$
\end{itemize} 

The average values are equal within the errors.
But the widths of the distributions differ significantly.
Thus, $\phi(NLS)/\phi(BLS)$ ratio has a maximum at $\log{(L_{bol})}=44.64$, which corresponds to $\log{(L_{[OIII]}/L_\odot)}=7.4$.
The dispersion of NLS distribution is lower, thus NLS occupy a narrower range in luminosity than BLS.

\section{The x-ray luminosity function}
 
We calculate a predicted soft x-ray(0.5--2kev) luminosity function based on the $L_{[OIII]}$ data.
[OIII]$\lambda$5007\AA~luminosities are converted into the soft x-ray following \cite{marconi04,heckman05}.
The result is plotted in fig.~\ref{fig2}\emph{(right)}.
For a comparison, predictions based on the $H_\alpha$ luminosity functions from \cite{schulze09,greeneho09} and [OIII] from \cite{haostrauss05} are also plotted in fig.~\ref{fig2}\emph{(right)}.
$H_\alpha$ luminosities are converted to the bolometric ones following \cite{greeneho07}, then from the bolometric ones to $L_{0.5-2kev}$ following \cite{marconi04}.
The x-ray LF of Sy1 based on the ROSAT, XMM-Newton and CHANDRA data \cite{hasinger05} is also shown in fig.~\ref{fig2}\emph{(right)}.
It should be stressed that our prediction corresponds to the actually observed x-ray luminosity function better than LFs from other studies. 
	 
\section{Results}
\begin{itemize}
 \item A method of evaluation of AGN luminosity function based on emission-line data accounting for variations of the density of galaxies due to the large-scale structure
 \item The [OIII]$\lambda$5007\AA emission-line luminosity functions for NLS, BLS, Sy1 (NLS+BLS) are obtained.
 \item At largest luminosities the LF of Sy1 agrees very well with the one of quasars \cite{croom04}.
 \item A prediction of soft x-ray luminosity function based on our emission-line LF is in an agreement with the observed x-ray LF.
 \item $\phi(NLS)/\phi(BLS)$ is not constant being a function of luminosity.
 It has a maximum between $\log{(L_{[OIII]}/L_\odot)}=6.8$ and $7.6$.
 This agrees with results of \cite{xukomossa12}.
 The average luminosities of NLS and BLS are equal within errors, but widths of the distributions differ.
 NLS occupy a narrower range of the AGN luminosity.
\end{itemize}

\end{document}